\documentstyle[epsbox,preprint,prl,aps]{revtex}
\tightenlines
\begin{document}
\draft
\title{Numerical Simulations on Szilard's Engine and 
Information Erasure}

\author{Takahiro Hatano and Shin-ichi Sasa}
\address{Department of Pure and Applied Sciences, 
University of Tokyo, 3-8-1 Komaba, Tokyo 153, Japan.}
\date{\today}
\maketitle

\begin{abstract}
We present a computational model for Szilard's engine and 
the information discarding process.
Taking advantage of a fact that the one is essentially 
the reversed cycle of the other, we can discuss the both by 
employing the same model.  Through numerical simulations 
we calculate the work extracted  by the engine and 
the heat generation in the information discarding process.
It is found that these quantities 
depend on some realistic ingredients, 
which means that the work done by the engine 
is no longer canceled by 
the heat generation in the information erasure.
\end{abstract} 


In 1876, Maxwell invented an external agent 
which utilizes information to extract the work from a single 
heat bath as a perpetual cycle \cite{maxwell}.
It is now known as Maxwell's demon.
One of the simplest configurations related to Maxwell's demon 
was presented by Szilard \cite{szilard}.
It consists of one molecule captured in a cylinder 
in contact with a heat bath. 
The demon inserts the piston in the middle of the cylinder, 
observes  which side the molecule is in,
and then expand it to extract the work 
from the heat bath. 
After the expansion, the demon removes the piston
and repeats the same manipulation.
Based on some assumptions, Szilard showed that 
the work of $k_B T \log 2$ can be extracted per one cycle. 
This cycle seems to violate the second law of thermodynamics 
and has provoked many arguments among 
physicists since presented \cite{book}.

Most of the physicists sought 
heat generation outside of the engine. 
In particular Brillouin \cite{brillouin} studied 
concrete measurement processes 
and attempted to prove that such measurements 
should be accompanied by  heat generation. 
Although his argument is attractive, one can always 
counterargue that there may be another measurement process 
without heat generation. 
Indeed Bennett suggested that reversible measurements are possible 
and proposed a new interpretation for the Maxwell's demon problem
\cite{bennett}. His argument is as follows. 
For the total system to be a complete cycle, 
the information of the preceding cycle stored in 
the manipulator must be discarded before succeeding cycles.
Following to the Landauer's claim,  logically irreversible processes 
such as information erasure should be accompanied by heat generation 
at least  $k_BT \log 2$ per bit \cite{landauer}.
It might be plausible that the extracted 
work of $k_BT \log 2$ by utilizing 1-bit information 
is compensated by heat generation of 1-bit information erasure.
Owing to their cancelation, the total system is expected 
to be consistent with the second law.

While the above argument seems reasonable, we have some doubts 
about the evaluation of the extracted work by the engine and 
generated heat in memory erasure, because the thermodynamic 
nature of the information might depend on its physical embodiment.
Especially the discussions so far are confined to 
idealized situation by means of thought experiments.
In this Letter, we study energetics of computational models 
for Szilard's engine and information erasure by
taking some realistic components into account.

\begin{figure}[htbp]
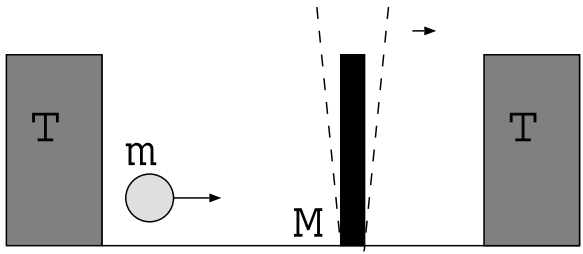

\begin{center}
\epsfile{file=model.eps}
\caption{Schematic figure of our model for Szilard's engine.}
\label{fig1}
\end{center}
\end{figure}

First, we present a computational model for Szilard's engine.
Our model is one dimensional so that we can describe this system 
by a velocity and a position of the piston $V,X$ and ones 
of the particle $v,x$. (See Fig. \ref{fig1}.)
Let the mass of the particle and the piston 
be $m$ and $M$, respectively.
The position of the particle is restricted to the region $-L<x<L$, 
where $L$ is the half length of the cylinder.
We assume evolution equations for $X$ and $x$ as 
\begin{eqnarray}
\label{piston}
M\frac{dV}{dt} &=& -\frac{\partial U(X,t)}{\partial X}
-\zeta V+\xi (t)+f(t),\\
\label{gas}
m\frac{dv}{dt}&=&-f(t)-g(t),
\end{eqnarray}
where $\xi (t)$ is Gaussian white noise whose statistical properties are
characterized by
$\langle \xi (t)\rangle =0$ and 
$\langle \xi (t)\xi(t')\rangle =2\zeta k_BT\delta (t-t')$.
$g(t)$ and $f(t)$ are the momentum transfer per unit time 
from the particle to the heat bath and to the piston,
respectively.

$g(t)$ is given implicitly by a stochastic rule at the
boundaries $x=\pm L$ in such a way that 
the particle is reflected being assigned new velocity $v$ at random with 
the probability distribution function 
\begin{equation}
\label{heatbath}
\phi (v)=\frac{m|v|}{k_BT}\exp [-\frac{mv^2}{2k_BT}].
\end{equation}
The distribution of the particle velocity turns out to be
Maxwellian when there is only a single particle in the system \cite{casati}.
Note that the piston is assumed not to undergo the reflection 
at the boundary.

The form of  $f(t)$ is given on the assumption that 
the piston and the particle collide elastically.
After a collision of the particle and the piston, they become 
\begin{eqnarray}
V' &=& \frac{1-\epsilon}{1+\epsilon}V 
+\frac{2\epsilon}{1+\epsilon}v,\\
v' &=& \frac{2}{1+\epsilon}V -\frac{1-\epsilon}{1+\epsilon}v,
\end{eqnarray}
where $\epsilon\equiv m/M$.
Since the momentum transfer from the particle to the piston is 
$2 m(v-V)/(1+\epsilon)$, we obtain
\begin{equation}
f(t)=\sum_i \frac{2 m}{1+\epsilon}(v-V)\delta (t-t_i),
\end{equation}
where $\delta (t)$ denotes Dirac's delta function and 
$t_i$ represents the time when $i$-th collision takes place.

\begin{figure}[htbp]
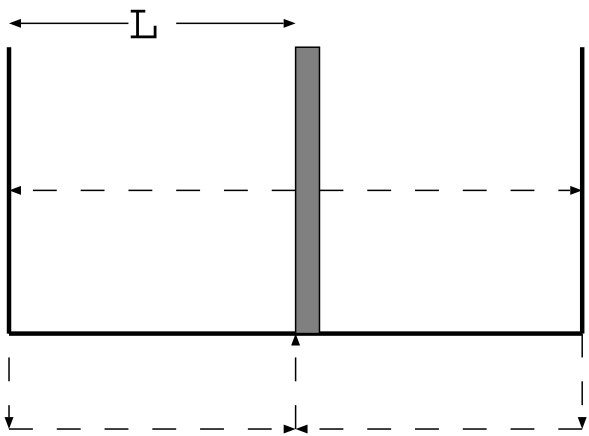

\begin{center}
\epsfile{file=manipulation_of_engine.eps}
\caption{Demon's manipulation of Szilard's engine.}
\label{fig2}
\end{center}
\end{figure}

The demon manipulates the piston through 
a trapping potential $U(x,t)$.
We assume the form of the potential as 
\begin{equation}
U(X,t)=\frac{k}{2}(X-X_0(t))^2.
\end{equation}
The form of $X_0(t)$ is given as the demon's manipulation.
In this paper, we assume 
\begin{equation}
X_0(t)=\left\{
\begin{array}{@{\,}ll}
\pm l t/\tau & \mbox{$ (0\le t\le\tau)$} \\
\pm l(2-t/\tau) & \mbox{$ (\tau\le t\le 2\tau)$},
\end{array}
\right.
\end{equation}
where the choice of the sign in the time interval  $0 \le t \le \tau$ 
depends on  the relative position of the particle to the piston. 
(The sign is positive for $X > x$, while negative for $x < X$.) 
The sign of $X_0(t)$ during $\tau \le t \le 2\tau$ is 
determined  so that $X_0(t)$ becomes continuous.
Further, the piston is 
assumed to  be removed at $t=\tau$ and to be reinserted
at $t=2\tau$. $f(t)$ becomes zero during $\tau\le t\le 2\tau$,
which means that the piston is outside of the cylinder.
In this way, the demon can repeat cycles.
The manipulation by the demon is shown in  Fig. \ref{fig2}. 

Note that $l$ can be larger value than $L$, because
the piston can collide with the particle even if $|X_0(t)|>L$.
Yet, since the difference of physical quantities such as work 
and heat between $l=L$ and $l>L$ is expected to be negligible 
when $k$ is large enough to localize the piston, 
hereafter we let $l=L$.

\begin{figure}[htbp]
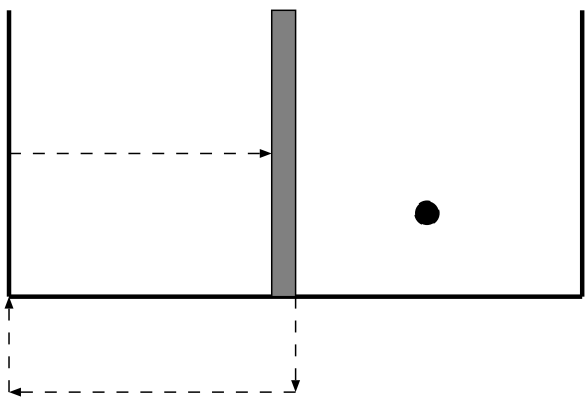

\begin{center}
\epsfile{file=manipulation_of_memory.eps}
\caption{Reversed operation of Szilard's engine.}
\label{fig3}
\end{center}
\end{figure}

Next we present  a computational model for a memory erasing process. 
We first notice that such a process 
can be designed as the reversed  one of a Szilard's cycle.
Initially, the piston is assumed to be in the middle of the cylinder.
The particle is confined in  one side (left or right) of it,
which  encodes an informational bit. After the piston is removed, 
it is reinserted in the left end  of the cylinder and moved to 
the middle. The particle is now in the right side.
This operation turns out to be the reversed one of a Szilard's cycle
as shown in Fig. \ref{fig3},  and  to be a logically irreversible process
to discard the information at the initial state (left or right)
as shown in Fig .\ref{fig4}.
Since we already have a computational model for Szilard cycles,
we can easily realize the reversed  Szilard cycle by employing 
the above model given  by  Eqs. (\ref{piston})  
and (\ref{gas}). All assumptions are the same except for 
the manipulation of the piston such that 
\begin{equation}
X_0(t)= \left\{
\begin{array}{@{\,}ll}
-L t/\tau & \mbox{$(0\le t\le\tau)$} \\
L(t/\tau-2) & \mbox{$(\tau\le t\le 2\tau)$}.
\end{array}
\right.
\end{equation} 
We also note that $f(t)=0$ when $0\le t\le \tau$.


We now discuss the energetics of our model.
We first assume that the removement and the reinsertion 
of the piston cost no energy. 
The validity of this assumption can be proved 
by analyzing a suitable model for this process
\cite{sekimoto2}. 
On this assumption, we study the energetics of
Eqs. (\ref{piston}) and (\ref{gas}).  
We follow the energetic interpretation for 
Langevin equations, which has been proposed 
by Sekimoto recently \cite{sekimoto}.
By multiplying Eqs. (\ref{piston}) and (\ref{gas}) 
by $V(t) dt$ and $v(t)dt$ 
respectively and integrating over one cycle, we obtain 
$$\int_0^{2\tau}MV(t)dV(t)
+\int_0^{2\tau}\frac{\partial U(X,t)}{\partial X}V(t)dt $$
\begin{equation}
\label{pistonenergy}
= \int_0^{2\tau} (-\zeta V(t)+\xi(t) + f(t))V(t)dt, 
\end{equation}
\begin{equation}
\label{gasenergy}
\int_0^{2\tau}mv(t)dv(t)=-\int_0^{\tau}f(t)v(t)dt
+\int_0^{2\tau}g(t)v(t)dt.
\end{equation}
These integrals are assumed as Stratonovich calculus 
for the following discussions being valid.
We analyze Eq.(\ref{pistonenergy}) first.
The first term of the left-handed side is written 
as the kinetic energy difference 
denoted by $\Delta K\equiv \Delta MV^2/2$. 
We can rewrite the second term as 
\begin{equation}
\int dU -\int \frac{\partial U(X,t)}{\partial t} dt
\equiv\Delta U +W, 
\end{equation}
where $W$ is defined as
\begin{equation}
W\equiv-\int_0^{2\tau} \frac{\partial U(X,t)}{\partial t} dt,
\end{equation}
which corresponds to  the work done by the engine.

The first two terms of the right-handed side of 
Eq. (\ref{pistonenergy}) is denoted by $-Q_1$, 
where $Q_1$ is interpreted to be the energy dissipation 
to the heat bath.
The last term then corresponds to the energy gain from the particle, 
which is denoted by $C$, that is,
\begin{equation}
C\equiv \int _0^{2\tau}f(t)V(t)dt.
\end{equation}
Then, Eq. (\ref{pistonenergy}) becomes 
\begin{equation}
\Delta K+\Delta U+W=-Q_1+C.
\end{equation}
Similarly, by analyzing Eq. (\ref{gasenergy}) as we did for 
Eq. (\ref{pistonenergy}), we obtain
\begin{equation}
\Delta K' = -C - Q_2,
\end{equation}
where $Q_2=-\int g(t)v(t)dt$ is the energy transfer 
from the particle to the heat bath and $\Delta K'$ is 
a kinetic energy increase of the particle.  
The total generated heat $Q$, 
the energy transferred from the system to the heat bath, 
is given by 
\begin{equation}
Q=Q_1+Q_2.
\end{equation}
Using these notations, we have en energy conservation law
\begin{equation}
\Delta K+\Delta K'+\Delta U+W+Q=0.
\end{equation}
Note that this expression holds for each of succeeding cycles.
By taking an average over many cycles, we obtain
\begin{equation}
\langle W \rangle+\langle Q \rangle=0.
\end{equation}
In the argument below,  $\langle \rangle$ denotes the average  
over many cycles.


Now we are ready to perform numerical simulations.
We calculated the time evolution by a second 
order Runge-Kutta method.
We let $k_BT=1, M=1,$ and $L=1$ for non-dimensionization 
and hence dimensionless parameters are $\epsilon$, $k$, 
$\zeta$ and $\tau$.
We are particularly concerned with the $\epsilon$ dependence 
of the work by the engine $\langle W\rangle _{e}$ and 
the heat generation in the memory $\langle Q\rangle _{m}$.
In Fig.\ref{fig5}, we plotted the result of simulations
with the parameter values $\tau = 10$, $k = 100$ and $\zeta = 0.1$.
For large $\epsilon$, the work and the heat generation 
goes below from $k_BT\log 2$.
It is also found that we get less work with smaller $k$.
Our simulations suggest that the maximum work $k_BT\log 2$ 
is obtainable in the limit of $\epsilon\rightarrow 0$, 
$k\rightarrow\infty$ and $\tau\rightarrow\infty$.

These results show that the compensation does not
occur unless we adopt exactly the same $\epsilon$ 
for the engine and the memory.
Moreover, total heat absorption in the engine cycle  
and the memory reset process is possible. 
Hence the interpretation of Maxwell's demon problem 
by Bennett and Landauer \cite{bennett} is not applicable to this model.

Furthermore, owing to the operationally inverse relation 
between the information erasure and the engine, 
we define the reversible heat generation $\langle Q \rangle_{rev}$ 
and irreversible one $\langle Q \rangle _{irr}$ as 
\begin{eqnarray}
\langle Q ^\rangle _{rev} &=& 
(\langle Q \rangle_{m}-\langle Q \rangle_{e})/2,\\
\langle Q\rangle_{irr} &=& 
(\langle Q \rangle_{m}+\langle Q \rangle_{e})/2.
\end{eqnarray}
In Fig. \ref{fig6}, $\langle Q \rangle_{rev}$ and 
$\langle Q \rangle_{irr}$ were plotted against $\tau$ 
while fixing the other parameter values as $\epsilon=10^{-4}$,
$\zeta=1$ and $k=100$. We found that $\langle Q \rangle_{rev}$ 
and $\langle Q \rangle_{irr}$ are closed to  $k_BT \log 2$ and 
 $2\zeta /\tau$, respectively. When the entropy production is 
defined through  the irreversible heat generation, it becomes zero 
in the quasi-static limit. Do not confuse this fact with an incorrect 
statement  that the heat generation during information erasure 
can be zero \cite{goto}, because the generated heat during logically 
irreversible processes has a positive reversible part.


Here we address three remarks on our results. First,
for generality of our results, we have studied another model 
where the single particle obeys the following equation 
instead of Eq. (\ref{gas}),
\begin{equation}
m \frac{dv}{dt}= -\bar{\zeta}\frac{dx}{dt} +\bar{\xi}(t)-f(t),
\end{equation}
where $\langle \bar{\xi}\rangle=0$ and $\langle 
\bar{\xi}(t)d\bar{\xi}(t')\rangle =2\bar{\zeta}k_BT\delta (t-t')$.
We confirmed that this model yields a qualitatively same graph as
Fig. \ref{fig5}.

Second, with some purturbative calculations, 
we get an analytic expression of the work done by the engine
\begin{equation}
\langle W\rangle_{e} =
\frac{1-\epsilon}{1+\epsilon}k_BT \log 2 -\frac{2\zeta }{\tau},
\end{equation}
where we have assumed that $\epsilon$ is small and 
$k \rightarrow \infty$. Similarly, as to the heat generation 
in the information erasure process, 
\begin{equation}
\langle Q\rangle_{m} =
\frac{1-\epsilon}{1+\epsilon}k_BT \log 2 +\frac{2\zeta }{\tau}.
\end{equation}
These expressions are good agreement with  the results of
simulations where $\epsilon < 0.03$.

Finally, we stress here the difference 
between our discussions and analysis recently presented  
by Magnasco \cite{magnasco}. 
His analysis is on the system described by a Fokker-Planck equation 
and applies to the automatic engine 
which needs no observer.
It was proposed by Popper \cite{popper} 
as an objection to the notion 
that information is equivalent to negentropy 
(negative entropy)\cite{brillouin}. 
While the engine needs no observer (hence no memory), 
Magnasco showed that it cannot work as a perpetual cycle.
His argument does not apply to the problem 
we discuss, since our system is assumed to be manipulated  
by the external agent which makes observation.


In conclusion, we invent a concrete model for Szilard's engine.
Numerical simulations show that its energy transformation ability 
from the heat to mechanical work depends on parameters, especially 
the mass ratio of the particle and the piston.
We also find that an information erasure process 
need not cost $k_BT\log 2$ energy in the same model.
As to the Maxwell's demon problem, 
the work obtained is not canceled with the generated heat
in the information discarding process.

The arising question out of our results is on the consistency 
with the second law. 
As mentioned above, the interpretation of the Maxwell's demon problem 
does not hold on the assumptions we adopt, 
and hence the second law neither.
As a plausible answer to the question, 
we conjecture that the excess heat is generated 
in the measurement process which transfers 
information from the engine to the memory.
Even if a reversible measurement is possible, 
it may be realized only for particular devices. 
(In our case, $\epsilon$ of the engine and the memory 
are precisely the same value.)
When the embodiment of the information in the memory part 
is different from that of the engine part, 
the excess heat may be necessary in the information 
transferring process.
These are future problems to be considered.

The authors acknowledge  T. Chawanya for constructive communication.
They thank M. O. Magnasco for his comments at an early stage
of this study. 
They also thank  K. Sekimoto, K. Kaneko and Y. Oono for 
discussions on related topics of nonequilibrium systems. 
This work was partly supported by grants 
from the Ministry of Education, Science, Sports and Culture of 
Japan, No. 09740305  
and from National Science Foundation, No. NSF-DMR-93-14938.


\begin{figure}[htbp]
\begin{center}
\epsfile{file=memory.eps}
\caption{Schematic figure of information discarding process.}
\label{fig4}
\end{center}
\end{figure}


\begin{figure}[htbp]
\begin{center}
\epsfile{file=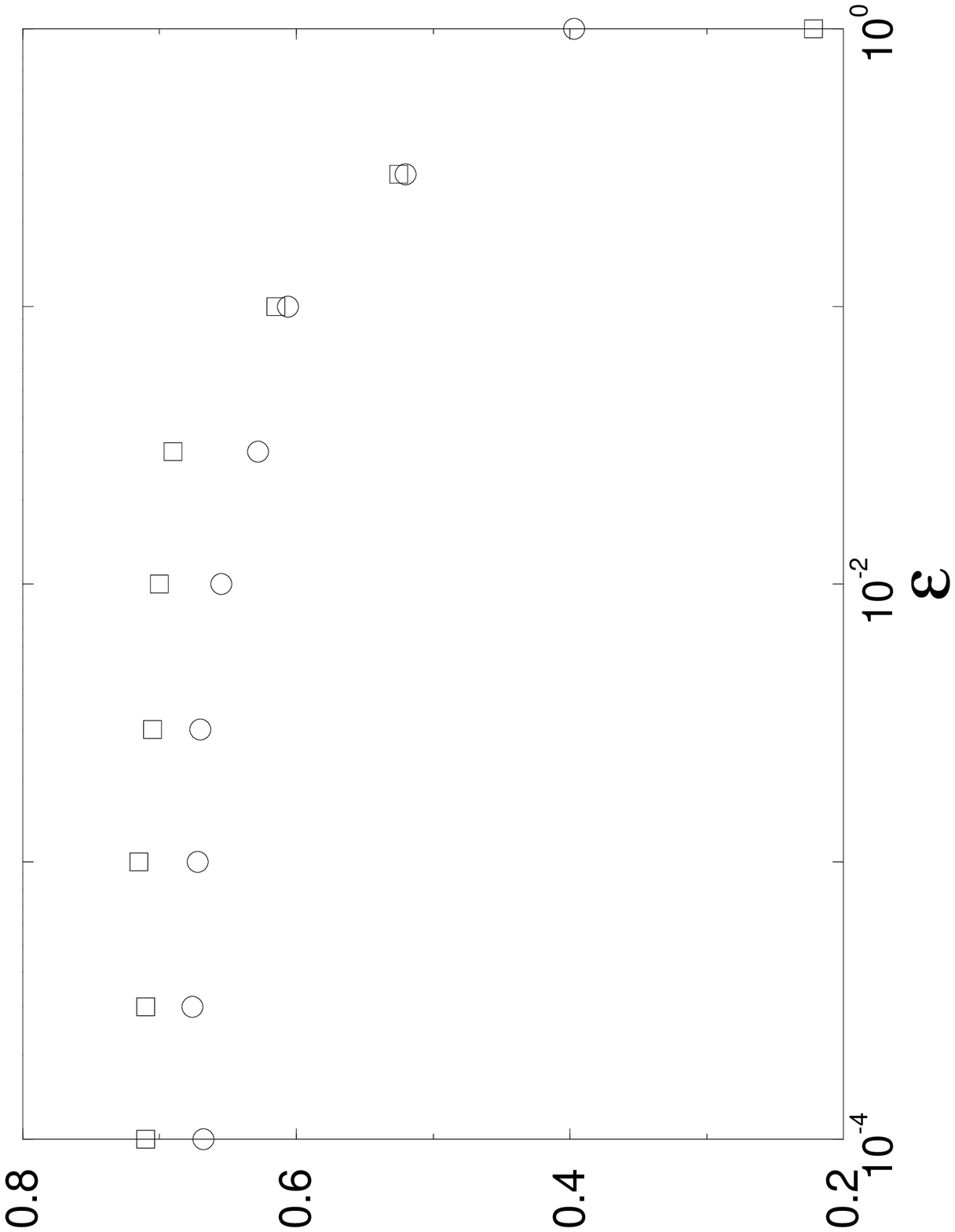}
\caption{$\langle W \rangle_{e}$ (circles) and 
$\langle Q \rangle_{m}$ (squares) versus $\epsilon$.  
These  were obtained as  averages  over 5000 cycles.}
\label{fig5}
\end{center}
\end{figure}


\begin{figure}[htbp]
\begin{center}
\epsfile{file=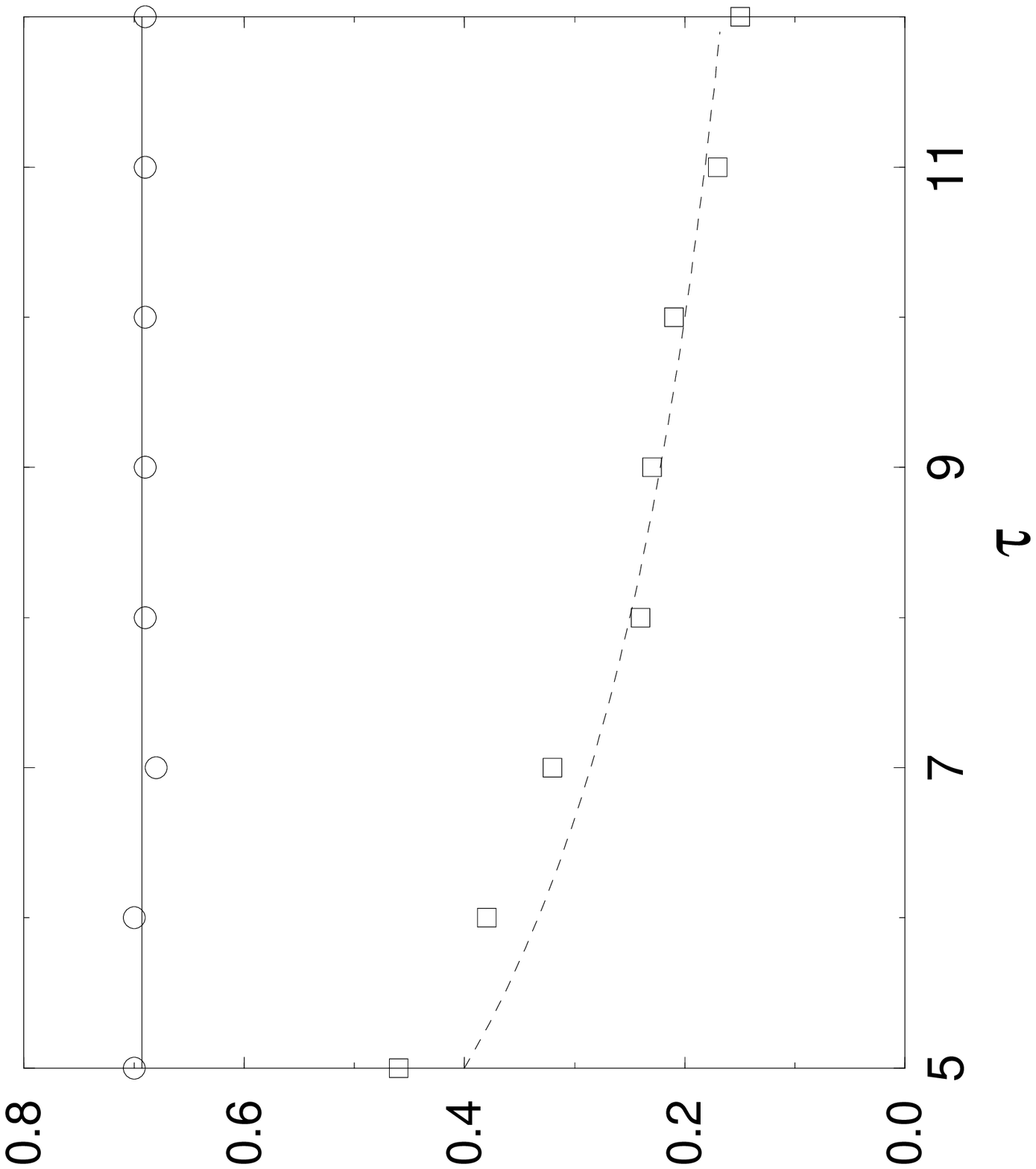}
\caption{$\langle Q \rangle_{rev}$ (circles) and
$\langle Q \rangle_{irr}$ (squares) versus $\tau$.
The solid line is $k_BT\log 2$ and the dotted line is $2\zeta/\tau$.}
\label{fig6}
\end{center}
\end{figure}

\end{document}